# PyFML- A Textual Language For Feature Modeling

A.F. Al Azzawi

Department of Software Engineering, IT Faculty, Philadelphia University, Amman.

## ABSTRACT

*The Feature model is a typical approach to capture variability in a software product line design and implementation. For that, most works automate feature model using a limited graphical notation represented by propositional logic and implemented by Prolog or Java programming languages. These works do not properly combine the extensions of classical feature models and do not provide scalability to implement large size problem issues. In this work, we propose a textual feature modeling language based on Python programming language (PyFML), that generalizes the classical feature models with instance feature cardinalities and attributes which be extended with highlight of replication and complex logical and mathematical cross-tree constraints. textX Meta-language is used for building PyFML to describe and organize feature model dependencies, and PyConstraint Problem Solver is used to implement feature model variability and its constraints validation. The work provides a textual human-readable language to represent feature model and maps the feature model descriptions directly into the object-oriented representation to be used by Constraint Problem Solver for computation. Furthermore, the proposed PyFML makes the notation of feature modeling more expressive to deal with complex software product line representations and using PyConstraint Problem Solver.*

## KEYWORDS

*Software Product Line, Feature Model, Python Programming language, Domain-Specific Language, Constraint Programming*

## 1. INTRODUCTION

A software product line (SPL) is considered as a set of products that all shared a lot of characteristics and vary in their particular configuration of variability, portfolios of very similar products are developed to get lower product cost, faster development cycles, and better quality [1], while Software product line engineering (SPLE) means to develop up a group of similar software products where commonality and variability among the family members (product variations) are explicitly determined by features where each feature is a visible characteristic of product in SPL related to product configuration[2]. Modeling variability is an essential in SPLE and a typical approach of the variability is called feature modeling and presents feature models as feature diagrams and the graphical notations is the most popular type of representation for feature diagrams [2], [3], particularly it encourages analysis to present a graphical language for these feature diagrams and associate it with formal representation considered as a base for building software engineering tool.

As far back as the first presentation of feature models [4], an extraordinary number of expansions has been made to the first presentation to address different requirements for example, having multiple clones of features (feature cardinality), including features in group relations (group cardinality)[5], [6], and adding attributes for feature [7]. In the same manner, graphical notation and textual notation are the two most main notations used to represent feature models by utilizing different subsets of propositional logic or Object Constraint Language (OCL) [8], but these expansions are not appropriately integrated into same feature modeling language or tool.

           41

In a survey of 37 works in feature modeling variability for SPL, 41% from these works support graphical notation, while 35% support textual notation, and the remaining do not give a sufficient details on the notations used, the textual notations are classified into three categories: similar to programming languages code, similar to XML, and embed representation in source code[9]. For graphical notation, it difficult to navigate FM graph especially in a large problem domain, and most of the works do not express feature attribute constructs and cross-tree constraints graphically well. For textual notations, it is difficult to read descriptions composed of programming code or XML, and it is not easy and hard to embed a representation.

We propose a textual feature modeling language (PyFML) that implements standard feature models with group relation, instance cardinality, feature attribute and complex cross-tree constraints. Moreover, the aim of this new language is to provide a textual human-readable notation that represents feature model, maps the feature model descriptions directly into the internal representation to be used by CPS Solver for computation and makes this notation of feature modeling more expressive and comprehensive to deal with complex software product line representations. Most of the works that automate feature model used either Prolog or Java programming language [8], [10], while Python programming language supports many features such as multiple paradigms, open-source, and large available packages and libraries which extends its capabilities with set theory, theorem provers, and predicate calculus. These features make it very suitable for scientific research [11], [12], and the highly desirable coding skill by recruiting companies [13].

For that, Python is well suited to host an embedding DSL's, since it enhanced with many tools that can be used to define, and implement a new language. Python is used to automate SPL's through building a feature modeling language (PyFML) and textX framework [14] is used to build a meta-model for the feature model so that the configuration can be easily transformed into CSP to automate feature model analysis and detect all valid configurations through the Constraint Problem Solver called "PyConstraint" [15], the results show that CSP technique can be used to implement feature model with thousands of features.tool.

## 2. BACKGROUND

In this section, we review main concepts of feature models, why we are using domain specification language rather than General-purpose language for developing a new textual feature modeling language, and then considering constraint satisfaction problem as a powerful paradigm for analyzing feature modeling constraint problems.

### 2.1. FEATURE MODELS

Feature model (FM) is represented by the hierarchical structure of related features with a root feature, and downward more detail levels using parent-child relations to define all valid configurations in a product line. Figure 1 (a) shows a basic case of a feature model representing different mobile phone, with variability in media and screen features [16]. A number of expansions have been made to this base representation to add an additional semantic such as feature cardinality, feature attributes, and cloning feature and in accordance to that, feature modeling can be divided into three main types: Basic models, Cardinality models, and Extended models.

For Basic models, parent-child relations are: feature is mandatory if it is required; feature is optional if it is optional; select feature leads to select its parent; select parent leads to select its children according to parent-child group type: And-group type leads to select all mandatory child features, Or-group type leads to select at least one child features, and alternative(xor)-group type leads to select exactly one child feature; Cross-tree constraints that relate features in different levels must be held (i.e. feature excludes another feature, feature requires another feature). A feature model can be represented by either graphical notation or textual notation as shown in figure 1.

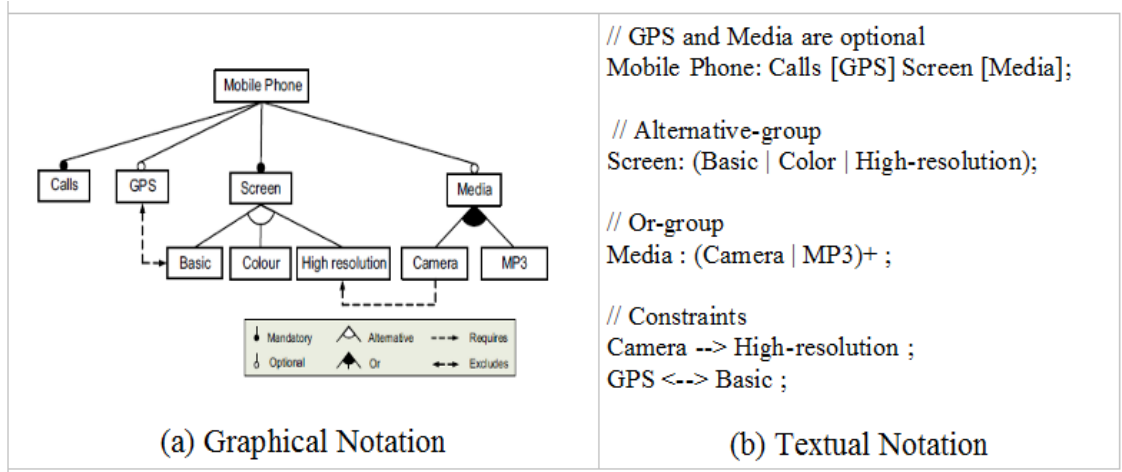

Figure 1. Feature Model Example

The cardinality of feature models is an extension of basic feature models with multiplicity [n..m] similar to UML class diagram multiplicity, that limits the number of child features from n to m if its parent is selected. This model is suitable when the product contains an arbitrary number of components [17]. While extended feature models have an extra information called functional attributes, where each attribute is composed of a name, a domain, and a value [16].

The propositional logic is most commonly used to represent the semantics of feature model to obtain all valid feature configurations where each feature considered as Boolean variable and each feature configuration must satisfy the propositional formulas corresponds to the feature diagram[18], [19]. The study in [20] compares between a graphical notation and a textual notation with respect of efficiency of analyzability and modifiability, and shows that the use of a textual notation is more efficient than graphical notation and in this work, a textual notation has been chosen first for its important and the alternative graphical notation that facilitates expressing and visualizing of feature model is left for future work.

## 2.2. DOMAIN SPECIFICATION LANGUAGES

General-purpose languages (GPL) are programming languages used for developing applications applicable in wide variety domains but lack the features of a specific application domain, while Domain-specific languages (DSL) are used to deal with the specific domain where a special syntax is built from scratch to implement applications within the specific domain [21]. Designing language using DSL for a specific domain inherent essential features from the domain perfectly, and manage these features through specification of restricted notations to solve the problems [21], [22]. Moreover, DSL has more expressive power than GPL, needs a short time to develop a correct application, and improves domain expert and programmer communication [23]. There are many DSL powerful tools towards DSL construction. Most popular representatives of these tools are Xtext, Spoofax, and MPS. These DSL tools are complex highly integrated with a specified development environment but also provide a powerful infrastructure for DSL development.
This work uses textX that inspired its grammar from Xtext with a lot of similarities; textX can be used in Python development environment and non-Python development using code generation from textX models, and its parsing doesn't need separate lexical phase while Xtext using definition rules to define lexemes and tightly integrates with Java[14], [24], [25].

## 2.3. CONSTRAINT SATISFACTION PROBLEM

Constraint programming (CP) is an effective paradigm for solving modeled constraint problems through searching techniques which may be guided by a set of algorithms or heuristics that deal with a constraint satisfaction problem (CSP). CSP comprises of a set of variables, a finite domain corresponding to each variable, and a set of constraints that restricts the values of these variables. CSP is solved by determines all values for variables in which all constraints are satisfied [26]. CP draws on a wide variety of techniques from computer science, artificial intelligence, and operations research and "provides a powerful reasoning capacity and better expressiveness" for feature modeling analysis [27].

There are many studies that used constraint programming to analysis feature models through mapping of FM to CSP and used a software solver to automate the analysis of the CSP. The mapping may vary depending on the CSP solver, but in general, the mapping is performed by considering each feature as a variable with the domain {0,1} or {true, false}, each relation in the FM is a constraint and an additional constraint may be added such that "assigning true to the root" [16]. However a limited of these attempts are presented experimental results and only a handful of analysis operations were implemented.

## 3. PYFML LANGUAGE FOR FEATURE MODELING

The textual specifications of FM is transformed to Python through model transformation technique, these transformations are performed in steps: i) build context free-grammar for the specification language and transform it to an equivalent textX grammar, ii) textX grammar is transformed to meta-model and generates dynamically Python classes accordingly, iii) these classes are updated to include new methods that perform FM analysis, and iv) integrate these classes with constraint solver PyConstraint These steps will be shown in the next section in more details.

PyFML is a textual DSL developed using Python programming language to automate feature modeling manipulation such as variability computation and constraint satisfaction. It allows the user to express feature modeling problems in a natural syntax that follows feature diagram and its constraints, rather than using general domain language to perform feature model analysis. PyFML combines textX Meta model framework, and Python-constraints solver with high-level features of Python such as lambda calculus and object-oriented design to satisfy semantic of the language. Figure 2 published in [28], shows the most possible choices to design a feature modeling language that automates FM analysis through feature model diagram, and PyFML design follows the shaded nodes in the graph

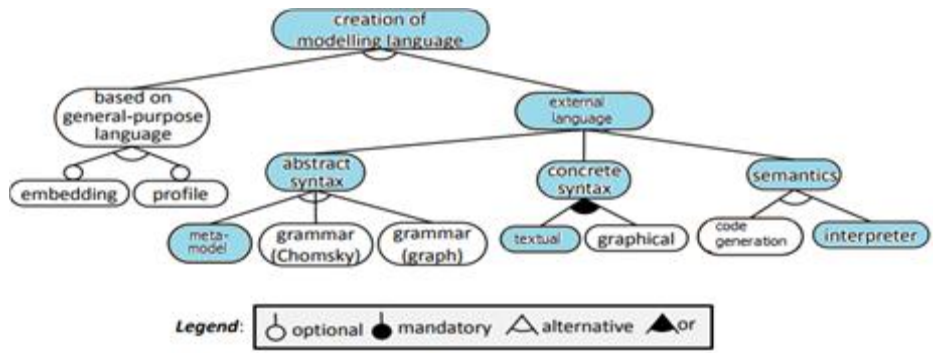

Figure 2. PyFML Design Choices

### 3.1. PYFML WORK FLOW

Figure 3 shows PyFML workflow that utilizes the textX framework to generate Python code. The EBNF of PyFML is transformed to textX grammar, than textX works as grammar interpreter which configured with a grammar to parse a textual input on the language specified by the grammar.

A meta-model is built from the configured grammar and that meta-model contains a Python class for each grammar rule. Besides, the textual input is parsed according to meta-model and converted to model, that model is a python object graph built from textual input and represents an Abstract Syntax Tree (AST), where each reference in AST is a python object reference for a class in meta-model and in our work, Figure 3 shows the PyFML workflow utilizing textX framework to generate Python code.

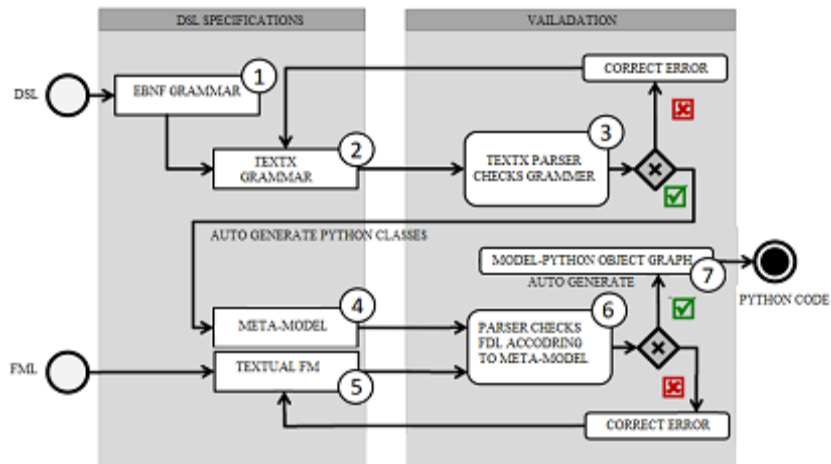

Figure 3. The PyFML workflow

### 3.2. PYFML GRAMMAR

The PyFML-feature modeling language is translated to Python code using the textX framework that hosts a parser called Parsing Expression Grammar (PEG). PEG is recursive-descent parser with many useful properties like limited backtracking, works in linear time, perfect parser and identical to Extended Backus-Naur Form (EBNF) [29]. The formal grammar descriptions of PyFML structures written in EBNF form is shown in Figure 4, the grammar describes atomic feature definitions, composite feature definitions, feature expressions and feature constraints.

```
FeatureModel= ID '=' CompoundFeature ";" | {Constraint};
Compound Feature=   Basic Feature [":" FeatureExpression ];
BasicFeature= ID [Cardinality] [":" "{" Attibutelist "}"];
Cardinality = "("  INT '.. .' INT ")";
Attibutelist=   Attribute | Attribute, ',', Attibutelist;
Attribute = ID "=" Value;
Value: INT | ID | STRING | FLOAT | BOOL;
FeatureExpression= Group "[" FeatureList "]" ;
Group= "all" | "one of" | "moreof";
Feature= BasicFeature| CompoundFeature;
FeatureList=   Feature | Feature, ',', FeatureList ;
Constraint= BooleanExp;
BooleanExp=FeatureName,LogicOp, FeatureName | "not", FeatureName;
LogicOp= "and"| "or"| "implies";
```

Figure 4. EBNF Grammar For PyFML Language

### 3.3. SYNTAX OF PYFML

The EBNF rules in figure 4 for PyMDL is converted to textX framework grammar rules in figure 5, these rules describe the syntactic structures of PyMDL using a concrete syntax and for each textX grammar rule, a Python class will be created at run-time and will be instantiate a concept during model parsing[25].

```
FeatureModel: fmname=ID '=' root=CompoundFeature ";"constraints*=BooleanExp;
CompoundFeature:  basicFeature=BasicFeature (":" featureexp=FeatureExpression)?;
BasicFeature : name=ID (cardinality=Cardinality)? (":" "{" attibutelist=Attibutelist "}")? ;
Cardinality: "(" start=INT '..' end=INT")";
Attibutelist: attibute+=Attribute [','] ;
Attibute: name=ID "=" value= Value;
Value: Val= INT | Val=ID | Val= STRING | Val= FLOAT | Val= BOOL;
FeatureExpression: group=Group "[" featurelist =FeatureList"]";
Group: name="all"|name="oneof"|name="moreof";
FeatureList: feature+=CompoundFeature [','];
BooleanExp: expression=Equal ";" ;
Equal: op=Greater ('==' op=Greater)?;
Greater: op=Less ('>' op=Less)?
Less: op=Implies ('<' op=Implies)?
Implies: op=Or ('implies' op=Or)*;
Or: op=And ('or' op=and)*;
And: op=Not ('and' op=Not)*;
Not: _not?='not' op=Operand;
Operand:  op=ID ('.' attr=ID)? |op=INT|op= FLOAT | op = STRING |op=BOOL| ('(' op=Equal ')');
```

Figure 5.  textX Grammar For PyFML Language

The following Figure shows how feature model in Figure 1 can be expressed using PyFML language.

```
FM = MobilePhone ( 1...1): all [
                    Calls   (1..1), GPS    (0...1) ,
                    Screen (0..1) : oneof[Basic (0..1), Color (0..1), HighResolution( 0..1)],
                    Media  (1..1) : moreof[Camera (0..1) , MP3( 0..1)]
       ];

Camera implies HighResolution;
Basic  implies  GPS;
GPS    implies  Basic;
```

Figure 6. PyFML language For FM in Figure 1

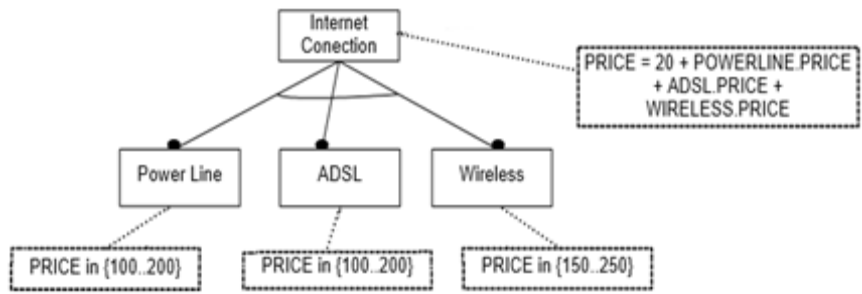

Figure 7. Part of attribute FM ( given in [16])

The following shows how attributed FM in above Figure 6 can be expressed using PyFML language.

```
FM = InternerConnection( 0..1) : {price=420}:
    oneof[
        PowerLine   (0..1): {price=150} ,
        ADSL    (0..1) : {price=150},
        Wireless (0..1) : {price=100}
      ];

InternerConnection.price=20 +ADSL.price+PowerLine.price +Wireless.price;
100<=PowerLine.price<=200;
100<=ADSL.price<=200;
150<= Wireless.price<=250;
```

Figure 8. PyFML Language For FM in Figure 7

### 3.4. META MODEL FOR PYFML

The framework textX is used to generate the meta-model from PyFML grammar rules, and a python class will be created dynamically for each class (grammar rule) in the Meta-model. Furthermore, the framework textX is used also to transform textual notation of the feature model written in PyFML language to a graph of python objects through meta-model instantiation.

Python object graph is actually a model structure or Abstract-Syntax Tree (AST), where each reference in the graph is resolved to an instance of a class in the meta-model [14]. Figure 8 shows the meta-model that characterizes the abstract syntax for PyFML grammar. The textual

International Journal of Software Engineering & Applications (IJSEA), Vol.9, No.1, January 2018

notation of the feature model, for example in Figure 6 is parsed according to the meta-model and the instantiated objects that are obtained at runtime according to meta-model classes shown in Figure 9.

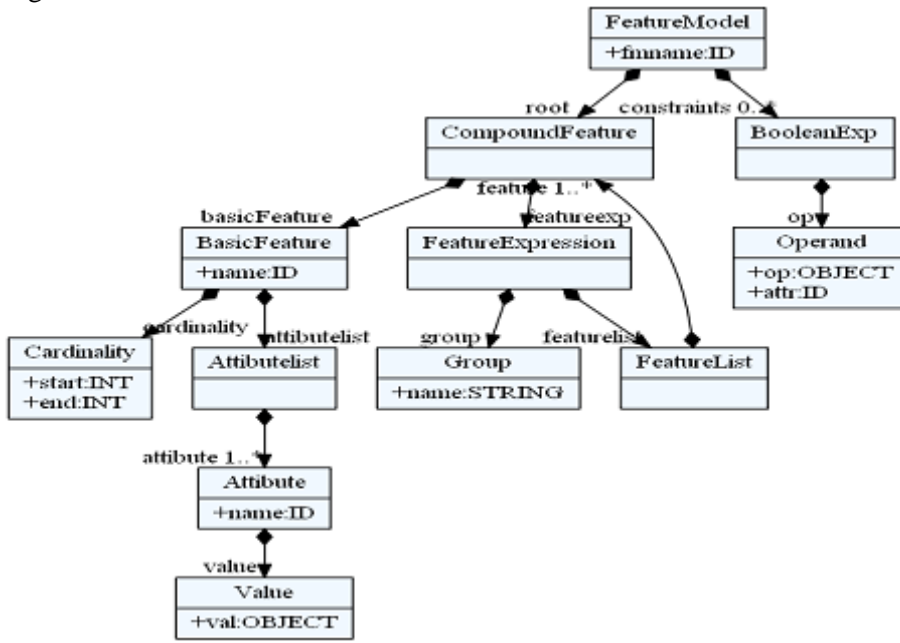

Figure 9. Meta-model of PyFML

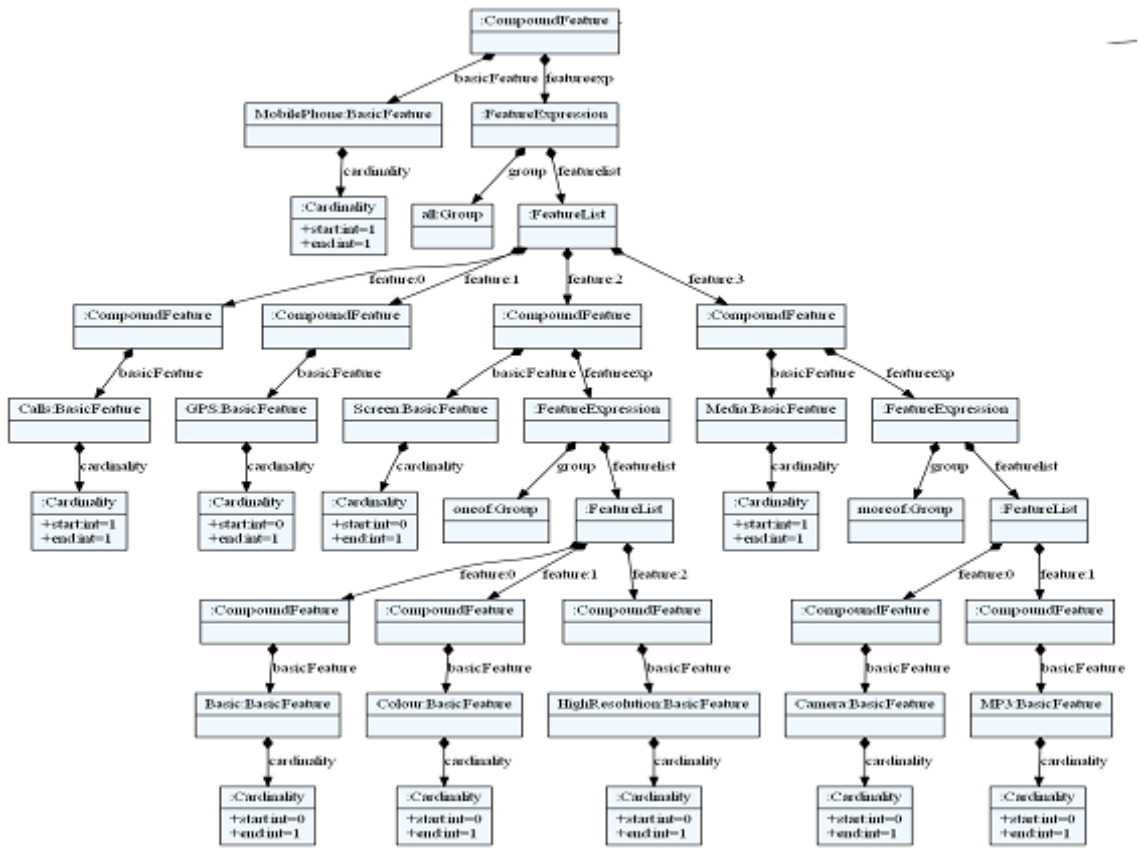

Figure 10. Python Object Diagram For Example 1

### 3.5. MAPPING FM TO CSP

FM deals with four main concepts: features, attributes, feature structure diagram, and cross-tree constraints.

**Features:** each feature is considered as a CSP variable with feature cardinality domain. Formally, if A is a feature with cardinality [m..n], then a CSP variable FV is created with domain [m..n], and the association of A with its clone is represented by FV $\Rightarrow$ v $\geq$ 0 where v $\in$ {m,n} and the following function is used to check the existing of a feature clone in a configuration

exists: FV$\rightarrow$ Bool where Bool = {T, F}:
exists(FV) = F iff FV = 0
exists(FV) = T iff FV > 0

**Attribute:** if A is an attribute for a feature F, then the attribute is represented as a CP variable with name F.A and the attribute domain.

**Feature structure diagram:** each group relation (and, or, alternative) is replaced by a logical expression to validate the connection between parent feature and its related child features, and Figure 10 shows how parent-child features relation is represented in diagram notation and its related CSP expression.

| Relation | Mandatory | Optional | Cardinality | Or | Alternative | Require | Excludes |
|---|---|---|---|---|---|---|---|
| Diagram Notation | A—•B | A—○B | A [n..m] B | A / B1 B2 BN \ | A / B1 B2 BN \ | A --> B | A <--> B |
| CSP EXPRESSION | (A>0) AND (B>0) | (A>0) AND (B>=0) | (A>0) AND (n<= B<=m) | (A>0) AND (B1>=0 OR B2>=0 OR BN>=0) | (A>0) AND (B1>=0 XOR B2>=0 XOR BN>=0) | (A>0) AND (B>0) | (A*B=0) AND (A>0) |

Figure 11. Diagram Notation and Related CSP Expression

**Cross Tree Constraints**: The cross-tree constraints in PyFML are expressed either by propositional formula to represent "requires" predicate, and "excludes" predicate or by Boolean operations.The constraints of FM are represented by the following mapping rules:

**Requires:** if A and B are features with cardinality domain D1= [m..n] and D2= [k..l] Respectively, where A requires B, then the constraint is equivalent to A $\Rightarrow$ B or (A>0 and B>0), which means that if A is chosen, then B must be chosen; according to the above existing function, A and B must have values greater than zero and in the domain D1, D2 respectively.

**Excludes:** if A and B are features with cardinality domain D1=[m..n] and D2=[k..l] respectively, where A excludes B, then the constraint is equivalent to (A* B=0 and A>0) or (A=0 and B=0), which means that if A is chosen, then B must be not chosen or both A and B not chosen.

**Boolean operations:** represents any other arbitrary relation between features or attributes and constraint relational constructs that are incorporated in CSP solver are: "or" , "and", "implies", "not", ">", ">=", "<", "<=" and "==" .

## 4. AUTOMATE ANALYSIS WITH PYCONSTRAINT

A feature model declaration together with its configuration are mapped into a PyConstraints problem, where PyConstraints is a constraint satisfaction solver used to define and implement python constraints. The constraints and the textX generated classes are combined along with decision procedures according to the structures of PyFML language. The FM CSP expressions shown in figure 9 are tested by PyConstraints for satisfiability. The PyConstraint problem for FM will be satisfiable when the configuration is valid for the provided FM declaration, and no constraints are violated; at that place, there is at least some possible product.

A PyConstraint model consists of a set of variables, a domain corresponding to each variable, and a set of constraints that restricts the values of these variables. Figure 11 shows a fundamental PyConstraint satisfiability problem, involving A and B features, and Mandatory relation between A and B. Lines 2–3 add two features as variables with domain [0,1], Line 4 assert the constraint to model Mandatory relation (A>0 and B>0) and running the problem suggests up to expectation it is satisfiable in the result.

```
1. problem = Problem()
2. problem.addVariable("A", [0,1])
3. problem.addVariable("B", [0,1])
4. problem.addConstraint(lambda a, b: a>0 and b>0, ("A", "B"))
5. print(problem.getSolutions() )

result :[{'A': 1, 'B': 1}]
```

Figure 12. PyConstraint Mandatory Problem.

The same procedure for automate analysis will be applied to other FM relations as in the above example except that the domain and the constraint must be changed according to FM relations shown in figure 9.

PyFDL language allows FM to have attributes, where each attribute will be exit when it assigns a value and we don't have to declare the type of these attributes, this will be handled internally in Python language since every attribute value considered as a datatype. Figure 12 shows FM PyConstraint problem, involving A and B features, where feature A has attribute p with values 0..10 and feature B has attribute m with value "ok" beside a constraint A.p>5 and B.m=="ok:. Lines 2–3 add the two attributes as variables with their values, Line 4 assert the constraint to FM PyConstraint problem and running the problem suggests up to expectation it is satiable in the result.

```
1. problem = Problem()
2. problem.addVariable("A.p", range(10))
3. problem.addVariable("B.m", ["ok"])
4. problem.addConstraint(lambda p, m:  p>5 and m=="ok",("A.p","B.m"))
5. print(problem.getSolutions() )
result :[{'A.p': 9, 'B.m': 'ok'}, {'A.p': 8, 'B.m': 'ok'}, {'A.p': 7, 'B.m': 'ok'}, {'A.p': 6, 'B.m': 'ok'}]
```

Figure 13. PyConstraint Attribute Feature Problem

## 5. INTEGRATING META MODEL AND PYCONTRAINT IN PYTHON

By using textX framework, every grammar rule is transformed to a Python class with a similar name dynamically; but these classes are insufficient to automate feature model analysis and are replaced with other classes that should be named with the similar name of the grammar rules so that can have extra methods to be integrated with PyConstraint. The following steps are considered to integrate meta-model with PyConstraint in Python:

1. Build a Python class for each grammar rule in the meta-model. These classes are instantiated during parsing of FM input (FM has written in PyFML language) to create graph objects that interpret by Python and PyConstraint to find all valid FM configurations

2. For basic feature class, a method is implemented that transforms attributes of the class which represent the basic feature to a variable and its corresponding domains, to be added to PyContraint problem.

3. For compound feature class, a method is implemented that transforms attributes of a class which represent feature and its relation (all, oneof, moreof) to avariable and it's domain, to be added to the PyContraint problem while feature relation is added as constraint to PyConstraint.

4. For attribute feature class, a method is implemented that adds attribute and its value as variable to FM PyContraint problem.

5. A set of classes are implemented for Boolean expressions (constraints), and in these classes, methods are implemented to evaluate the Boolean expressions.

6. For constraint class, a method is implemented to add a constraint to FM PyConstraint problem and to call a Boolean evaluation method.

The complete code that implements PyMDL, can be found at the following link https://drive.google.com/file/d/1TmxB4fR4oTeBgk4Z6vbfr0R5xEkMUrJ7/view?usp=sharing

## 6. CONCLUSIONS AND FUTURE WORKS

This work automates FM analysis based on mapping of FM into CSP solvers and suggests a domain-specific language called PyFML for feature modeling. This language supports feature cardinality, feature attributes, group relations, multiple groups, and complex constraints, and allows the user to express feature model and its constraints in a natural readable language, rather than restrictive standard syntax form required solvers or programming language. PyFML easily transforms textual notation of feature paradigm and constraint solver to analysis feature modeling.

We use model transformations approach to transform textual representation of FD to meta-model than an object-oriented representation is generated in Python programming and PyConstraint solver is used to validating products according to generated model dependencies and its constraints. PyFML language can easliy be extended or updated to support new other characteristics, just by changing the suggested grammar while Python programming language as development environment can enhance these changes with presence of many tools that focus on theorem proves, predicate calculus, set theory, and algebraic expression to implement these changes, and to optimize the new languages. This work can be extended to provide SPL with a variability analysis of different configurations supported by product line and generate the graphical representation for the textual representation.